\documentclass[nofootinbib,superscriptaddress,onecolumn,preprintnumbers]{revtex4}
\usepackage{graphicx}
\usepackage{amsmath,amssymb}
\usepackage{color}

\newcommand{\reff}[1]{(\ref{#1})}
\newcommand{\eref}[1]{Eq.\reff{#1}}
\newcommand{\erefs}[1]{Eqs.\reff{#1}}
\newcommand{\figref}[1]{FIG.\ref{#1}}

\begin{document}
\title{Analysis of non-diffusive avalanche transport of energetic particles}

\author{N. Carlevaro}
\email{nakia.carlevaro@enea.it}
\affiliation{Nuclear Department, ENEA C.R. Frascati, Via E. Fermi 45,  (00044) Frascati, Italy}

\author{M.V. Falessi}
\affiliation{Nuclear Department, ENEA C.R. Frascati, Via E. Fermi 45,  (00044) Frascati, Italy}
\affiliation{Istituto Nazionale di Fisica Nucleare (INFN), Sezione di Roma, P.le Aldo Moro 2, (00185) Roma, Italy}

\author{G. Montani}
\affiliation{Nuclear Department, ENEA C.R. Frascati, Via E. Fermi 45,  (00044) Frascati, Italy}
\affiliation{Physics Department, ``Sapienza'' University of Rome,  P.le Aldo Moro 5, (00185) Roma, Italy}

\author{Ph. Lauber}
\affiliation{Max Planck Institute for Plasma Physics, Boltzmannstrasse 2, D-85748 Garching, Germany}

\begin{abstract}
The dynamics of energetic particles (EPs) interacting with Alfvén eigenmodes (AEs) for the ITER 15MA baseline scenario was described using a reduced 1D model in \cite{carlearo-ppcf}, and successfully tested against nonlinear wave-particle simulations. In this paper, we introduce a detailed phase-space and statistical analysis of this case to characterize the emerging EP transport regimes. Deviations from pure diffusive dynamics are quantitatively addressed, indicating the limitations of standard quasi-linear descriptions. The phase space diagnostics introduced allows to describe the emergence of a very complex dynamics of overlapping resonances and substructure formation, reinforcing the evidence of non-diffusive domino-like AEs excitation.
\end{abstract}

\maketitle

\section{Introduction}

The achievement of controlled thermonuclear fusion remains one of the major scientific and technological challenges of modern physics. Reaching this objective requires bringing the plasma to supra-critical conditions in which fusion reactions become self-sustaining. Two main strategies have been developed to access this regime. The first is inertial fusion (IF) \cite{IF1zhou,RT-zhou,RT-zhou-book}, where intense laser pulses compress and heat a fuel target to ignition conditions. Recent experiments \cite{Abu-Shawareb:24} reported a target energy gain of about 1.5, exceeding the Lawson criterion \cite{Lawson:57} for self-sustained burn. The second approach is magnetic confinement fusion (MCF), which relies on magnetic fields to confine high-temperature plasmas for times sufficiently long to allow significant fusion reactions, and is widely regarded as one of the most promising routes toward a future fusion reactor \cite{Gibney:22}. Although both approaches aim at achieving fusion ignition, they rely on fundamentally different confinement mechanisms: laser-driven implosion in IF and magnetic confinement in MCF, therefore involving distinct physical and technological challenges. In IF, the efficiency of the laser-to-plasma energy coupling is a critical issue. In contrast, MCF performance is largely limited by turbulent transport processes \cite{Ham:20}.

In this respect, one of the primary challenges in MCF is being able to predict the dynamics of a burning plasma and describing $\alpha$-particle and, more in general, energetic particle (EP) dynamics, including supra-thermal ions and electrons generated by external heating, is crucial  \citep{HW20-review,ZCrmp,ZC15njp,dz13,bass10,chapter7,pucellaFTU}. It is important to remark how, when addressing the physics of EP and, in particular, the EP-driven instabilities, the gyro-kinetic theory \citep{ZCrmp} should be implemented for providing valuable insight into the fundamental physics processes. In fact, EP transport processes involve resonantly excited fluctuations, which exhibit different time scales and structures compared to thermal plasma instabilities. Furthermore, EPs may excite singular radial mode structures at shear Alfvén wave continuum resonances which, in turns, can generate, by mode conversion, radially propagating microscopic fluctuations absorbed at different radial positions \cite{zonca-ijmpa,ZC15ppcf,falcon1,falcon2}. Nonetheless these relevant aspects can be captured by global gyro-kinetic simulations, such approaches are, in turn, extremely resource-intensive and time-consuming when applied to both current and next generation devices \citep{bierwage18,La13,DTT_2021}. Consequently, simplified reduced models describing EP dynamics are of crucial importance. One of the key challenges of these approaches is accurately identifying and incorporating transitions between different EP transport regimes (due to the intermittent nature of EP dynamics \citep{duarte2019nf}). 

The 1D bump-on-tail (BoT) paradigm has often been used to mimic the nonlinear interplay between EPs and Alfvénic fluctuations \citep{BB90a,BB90b,BB90c,CZ13,ZC15njp} (see also \cite{KV20,ARKN20,KV14}, for application of the Hamiltonian formulation of the BoT in other context of plasma physics). Details of this paradigm can be found in \cite{jpp_cmf}, where traditional studies are refined with new numerical simulations highlighting the characteristics of phase-space trapping dynamics in nonlinear regimes for both single and three resonance cases. Examining the role of the nonlinear velocity spread in mode overlap (i.e. the velocity variation due to the exchange of energy with the resonant mode), unlike predictions based solely on the size of rotating trapping clump, an additional factor is found necessary. This is related to the effective deformation of the velocity distribution function, allowing a more accurate depiction of resonance regions. It is also shown how overlapping resonances enhance fluctuation levels, increasing phase-space energy transfer to Langmuir modes. Moreover, we remark how the BoT paradigm constitutes a general framework of investigation, impacting different fields of plasma and non-linear physics. In fact, an important isomorphism exist between the beam-plasma interaction and the working mechanism of a Free Electron Laser \cite{Duccio05,CFGGMP14}. Thus, the technical tools here employed has to be regarded as a more general achievement with respect to the direct interest for the Tokamak configuration, providing physical insight in a number of relevant plasma physics contexts. 

In \cite{spb16}, an analysis of the ITER 15MA baseline scenario suggests that realistic nonlinear simulations are not accurately characterized by QL model, as it fails to predict the proper \textit{domino}-like mode excitation \citep{BB95b,2020PhPl27b2117W,2019PhPl26l0701D}. In particular, modes can be excited due to resonance overlap, since the resonant particle trapping and rotation results in an enhanced drive by creating steeper phase-space gradients at the radial location of the adjacent modes. This, in turn, generates an avalanche transport of EPs until the last mode is saturated. This is found to be related to an enhanced saturation of the sub-dominant part of the spectrum. Although the excitation phenomenon of stable modes close to marginal stability can be accurately described by QL scheme (see the pioneering paper \cite{ivanov67}), the damping rate radial profiles of this scenario bring to a peculiar situation where the predictability of the QL theory is reduced. Such a mechanism was instead well captured by the 1D reduced description introduced in \cite{carlearo-ppcf}. A mapping procedure linking the radial coordinate of the realistic 3D scenario to the velocity dimension of the BoT paradigm, described by a Hamiltonian beam-plasma model \citep{ZCrmp,CFMZJPP,ncentropy}, has been defined and successfully tested against EP profile evolution predictions from HAGIS-LIGKA wave-particle nonlinear simulations \citep{lauber2007ligka,pinches1998hagis}.

In this paper, we provide a detailed characterization of the transport associated to the domino mode excitation mentioned above, in the presence of the least damped 27 toroidal AEs (TAEs), as addressed in \cite{spb16}. In order to analyze the transport phenomenology in the phase space, we implement the Lagrangian Coherent Structure (LCS) technique \citep{Haller_2015}. LCSs divide the particle phase-space into macro-regions where fast mixing transport processes occur and, for this reason, they are routinely used to describe transport processes in a wide range of contexts and, specifically, to plasma physics in \cite{MPJPP,CFMZJPP,Di_Giannatale_2018, Di_Giannatale_2018b,Pegoraro_2019, padberg2007lagrangian, rempel2023lagrangian, Bo11a,2021NucFu..61g6013D,2023PPCF...65g4001W}. While this technique was firstly implemented in \cite{jpp_cmf} for the BoT paradigm in presence of one and three interacting resonances, here we generalize this approach to the realistic scenario of \cite{spb16}, described by the means of the mapping procedure. The use of this methodology provides an immediate representation of the mechanism behind the avalanche transport. In particular, the formation of the stiff gradient of the EP distribution function at the boundary of the plateau region drives the development of elongated LCS and, consistently, enhances transport while exciting at the same time the sub-dominant part of the spectrum through nonlinear wave-particle interactions. This scenario differs significantly from an artificial constructed benchmark case, in which the whole spectrum is linearly globally unstable. The evolution of the distribution function is then analyzed by tracking the mean square displacement (MSD) of test particles over time. We show how particles related to the avalanche response exhibit non-diffusive (convective) transport. This indicates that, when stable modes are destabilized in this specific setup, a complex transport pattern is generated that may require a fully-nonlinear treatment of EP dynamics. An additional evidence of the non-diffusivity of transport is provided by the analysis of phase-space fluxes. This work thus provides a quantitative description of the transport traced in \cite{spb16} by using a reduced model. Due to the small computational cost, this tool can be used to obtain indicators of specific transport phenomena that can guide the realistic 3D simulations.

We conclude this introductory section by discussing the important role played by sources and sinks. Fast ions are continuously replaced by incoming fluxes, preserving the steady profile with respect the losses due to outward transport. Taking into account these two additional phenomena is certainly relevant when we are interested to the evolution of a steady EP profile. When analyzing the ITER 15MA baseline scenario, the objective of \cite{spb16} is the demonstration of the physical processes underlying the relaxation mechanism of EP. In this sense, the presence of sources and sinks (and wave-wave nonlinearities) is not taken into consideration: the nonlinear dynamics has indeed a very fast time scale compared with the corresponding plasma heating and slowing-down time, which are responsible for the steady state configuration, i.e. a sort of equilibrium between source and sink fluxes \citep{ZCrmp}. This approximation, well grounded for a physics oriented analysis, is maintained also in the present work. In fact, in view of properly reproducing by means of reduced modeling the distribution function relaxation of the HAGIS-LIGKA simulations discussed above, we also disregard additional inward and outward EP fluxes.  Notably, when studying a collisionless framework, the saturation mechanism consists of the depletion of the free energy of the system through the flattening of the distribution function. In this sense, the analysis of the phase mixing of EP and the related transport results clearer in this scheme. Nevertheless, it should remain clear the importance of sources and sinks when investigating their influence of EP relaxation over a sufficiently long time scale. This will be addressed in future works by studying the role of dissipation in governing the nonlinear dynamics and saturation level of mode instabilities, when continuous injection of EP is addressed.

The paper is organized as follows. In Sec. \ref{sec2}, the 1D reduced model is introduced and applied to the EP dynamics of the ITER 15MA baseline scenario. A benchmark case is also discussed. In Sec. \ref{sec4}, the LCS technique is applied to characterize the dynamics in the phase space and the transport barriers of both systems. In Sec. \ref{sec3}, a statistical analysis of tracer dynamics and the study of the fluxes are addresses to define the nature of the underlying transport processes.

\section{1D reduced model for AE driven turbulence}\label{sec2}
In \cite{carlearo-ppcf}, the BoT model is used as a reduced framework to quantitatively describe the dynamics of EPs interacting with AEs. In this section, we briefly review the main elements and key results, with the aim of providing the necessary context for the present analysis. In particular, we summarize the essential features of the reduced model and its application to the 15MA ITER baseline scenario. This framework is extensively studied in \cite{spb16}, where the HAGIS-LIGKA hybrid simulations are used to analyze the radial EP redistribution when multiple TAEs are considered. It is outlined the importance of the linearly subdominant modes and that the EP spread exceeds the QL. 

The dynamic of EP/TAE interaction is modeled implementing a linear one-to-one correspondence between the normalized small radius $s$ (thus $s\in[0,1]$) of the Tokamak configuration, and the velocity $u$ of the BoT model. The latter 1D scheme considers a fast charged particle beam (of density $n_B$) as a bump in the tail of a Maxwellian equilibrium configuration in the velocity space. In this sense, it can mimic the behavior of EP. This beam interacts with the plasma considered as a cold dielectric medium of density $n_p$, which supports longitudinal electrostatic (Langmuir) waves. By virtue of the mapping between the two systems, the reduced representation of the AEs corresponds to the scalar electrostatic modes $\phi_j(\tau)$ of the BoT model with mode number $\ell_j$, having frequency $\omega_j$, linear drive $\gamma_{Dj}$, damping rate $\gamma_{dj}$ (the effective mode growth rates are thus $\gamma_j=\gamma_{Dj}-\gamma_{dj}$). The resonance condition writes as $\ell_j=1/u_{rj}$, where $u_{rj}$ denotes the resonant velocity). Here and in the following, all quantities are dimensionless (in particular, frequencies and the time $\tau$ are normalized using the plasma frequency), and we use capital letter, i.e. $\Omega_j$, $\Gamma_{Dj}$, $\Gamma_{dj}$ and $\Gamma_{j}$ for the AE counterpart parameters. 

The linear correspondence between $u$ and $s$ is derived locally for a single resonance using a map of the resonance condition, i.e. $\Omega_j-\Omega(s)\propto\ell_j(u-u_{rj})$ \citep{carlearo-ppcf}. We assume this relation to be valid in the whole radial domain and the linear link ultimately reads $u=(1-s)/\ell^*$, where $\ell^*$ is an arbitrary constant defining the spectral features. In fact, mode numbers are integers set now by the conditions $\ell_j=\ell^*/(1-s_{rj})$, where $s_{rj}$ denotes the AE resonant radii, and $\ell^*$ can be arbitrarily fixed in order to have enough distinct integers $\ell_j$ to properly represent the AE spectral density. 

In this scheme, the nonlinear self-consistent equations for EP relaxation are thus equivalent to the BoT dynamical system. In the present analysis, we adopt the Hamiltonian $N$-body formulation of the BoT model \citep{CFMZJPP,ncentropy}. We remark that the reduced phase space describing wave-particle trapping phenomena is defined by $(x,u)$, where $x$ represents the spatial dimension of the BoT model (the 1D plasma is considered as a slab of periodicity $2\pi$). In order to provide a direct comparison with the HAGIS-LIGKA simulations and to discuss the relevant features of the EP redistribution toward the edge of a Tokamak configuration, in this work we adopt directly the quantity $s$ in place of $u$ and consider it as a ``velocity'' variable defining the phase space $(x,s)$. Thus, indicating by $(x_i,s_i)$ a single particle position of the $N$-body scheme, the considered dynamical system writes as
\begin{subequations}\label{mainsys1}
\begin{align}
&\dot{x}_i=(1-s_i)/\ell^* \;,\label{eq1}\\
&\dot{s}_i=-\ell^*\sum_{j=1}^{M}\big(i\,\ell_j\;\phi_j\;e^{i\ell_jx_{i}}+c.c.\big)\;,\label{eq2}\\[-10.5pt]
&\dot{\phi}_j=-i\phi_j-\gamma_{dj}\phi_j+\frac{i\eta}{2\ell_j^2 N}\sum_{i=1}^{N} e^{-i\ell_j x_{i}}\;,
\end{align}
\end{subequations}
where the dot denotes differentiation with respect to the normalized time $\tau$ and $N$ and $M$ indicates the total number of particles and modes, respectively. Here, we have introduced the parameter $\eta=n_B/n_p$, which relevance will be detailed below. The linear stability/instability of the modes $\phi_j(\tau)$ is governed by the following linear dispersion relation:
\begin{align}\label{disrel}
2(\omega_j+i\gamma_{Dj}-1)+\frac{\eta\ell^*}{R\ell_j}
\int\!\!ds\frac{\partial_s f_{H}^0(s)}{\ell_j(1-s)/\ell^* - \omega_j-i\gamma_{Dj}}=0\;,
\end{align}
where the integral is defined over the small radius domain and $R=\int ds f_H^0$ (boundary terms should be taken into account in the numerical integration). The typical initial particle distribution is represented by a slowing down profile
\begin{align}\label{fhot}
f_H(\tau=0,s)=f_H^0(s)=B_1\,\textrm{Erfc}[B_2+s B_3]\;,
\end{align}
where $B_{1}, B_{2}, B_{3}$ are given constants.

We remark here that, due to the reduced dimensionality of the model (from 3D to 1D), $\gamma_{Dj}$ result smaller than the normalized linear AE drives $\Gamma_{Dj}$, and their ratio roughly corresponds to the fraction of the most resonant particles for each resonance \citep{carlearo-ppcf,nceps22}. Moreover, the damping are defined by scaling the AE damping $\Gamma_{dj}$ (provided by the background plasma and not by the resonant EP \citep{Ph15}), i.e. $\gamma_{dj}\equiv\Gamma_{dj}\omega_j/\Omega_j$. Eq.\reff{disrel} is therefore crucial for the closure of the mapping: once the parameter $\eta$ is fixed, the dispersion relation provides, as a result, the linear quantities $\omega_j$ (which in turn defines $\gamma_{dj}$) and $\gamma_{Dj}$. In this work, we adopt the same empirical procedure introduced in \cite{carlearo-ppcf}. As pointed out in \cite{spb16}, the fully self-consistent EP redistribution immediately after saturation closely matches the QL predicted profiles, while only at later times does an avalanche-like transport toward the edge occur. We thus determine $\eta$ by performing a parameter scan to match the relaxed flattening width predicted by the QL approximation when applied to both schemes, obtained through direct integration of the PDE system of the QL model. This constitutes the closure scheme adopted in the present analysis, but, in principle, the QL flattening of the physical EP/AE system can already be estimated from linear quantities through the standard scaling $\Delta s\propto \Gamma_{D}^2$ \cite{spb16}. The dependence of the system on the parameter $\eta$ is particularly significant, as it ultimately determines the strength of the linear drive and, consequently, the width of the plateau region. As will be shown later, the effects associated with reducing the drive reflect the dimensional reduction inherent in the BoT model with respect to the full EP/AE system. This is related to the fact that, in a realistic 3D configuration, not all particles are resonant. Therefore, a reduction of the effective drive is required in the reduced model in order to mimic the proper EP redistribution. It is also important to remark that, when multiple modes are included in the dynamics, the mapping procedure is implemented by specifying a single reference resonance (with average characteristics). In this way, all the ingredients of the BoT paradigm can be defined without ambiguity (especially the drive parameter $\eta$). This intrinsic discrepancy between the original and the mapped system could be overcome by adapting the mapping to each resonance through a sort of ``patch map'' that include also an estimate of the portion of the most resonant particle in the 3D system. Nonetheless, in the case of sufficiently overlapping resonances (as in the present scenario), the error introduced by describing all resonances within the single reference framework results in a negligible effect \cite{carlearo-ppcf}.

\begin{figure}
\centering
\includegraphics[width=0.45\textwidth]{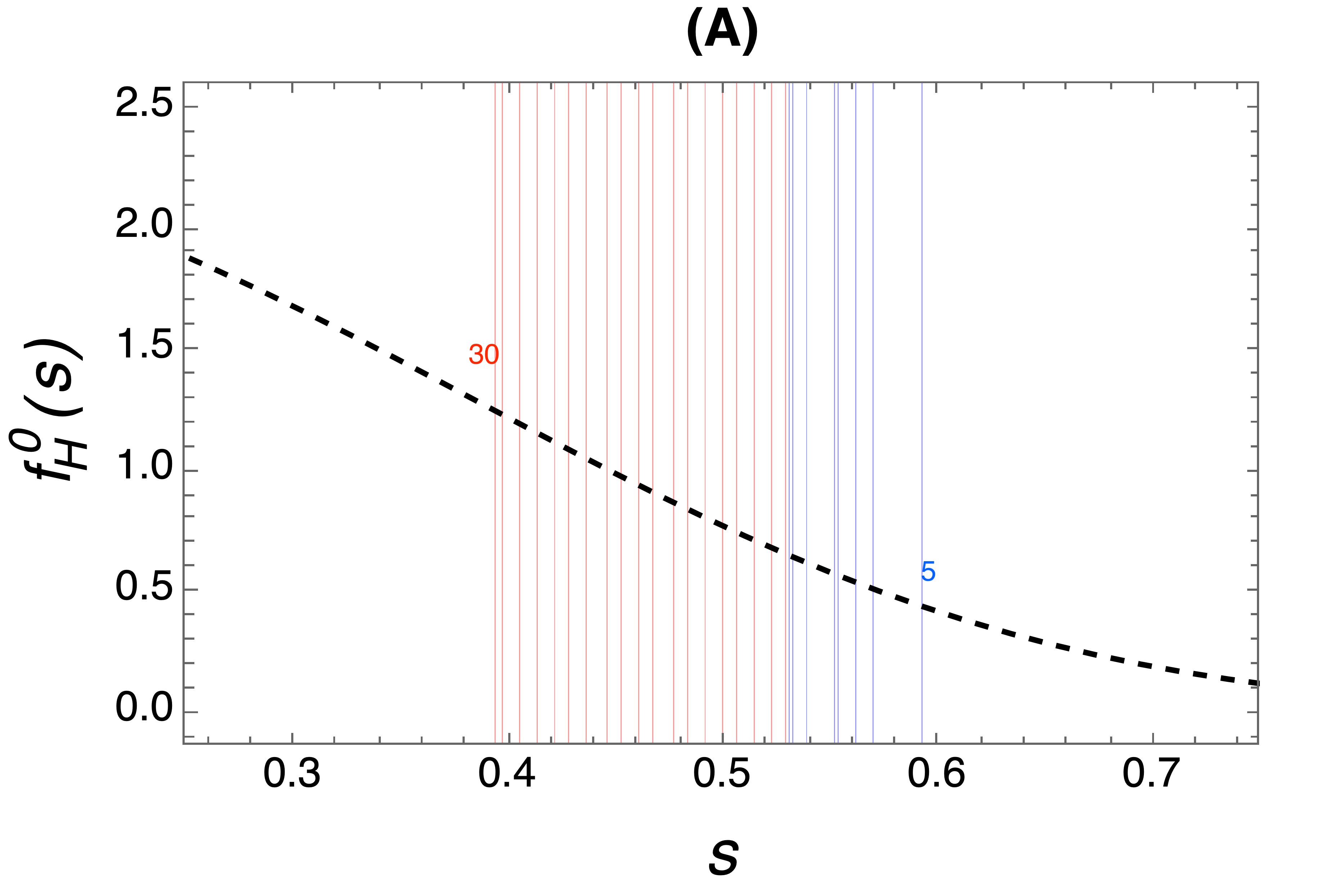}\;
\includegraphics[width=0.45\textwidth]{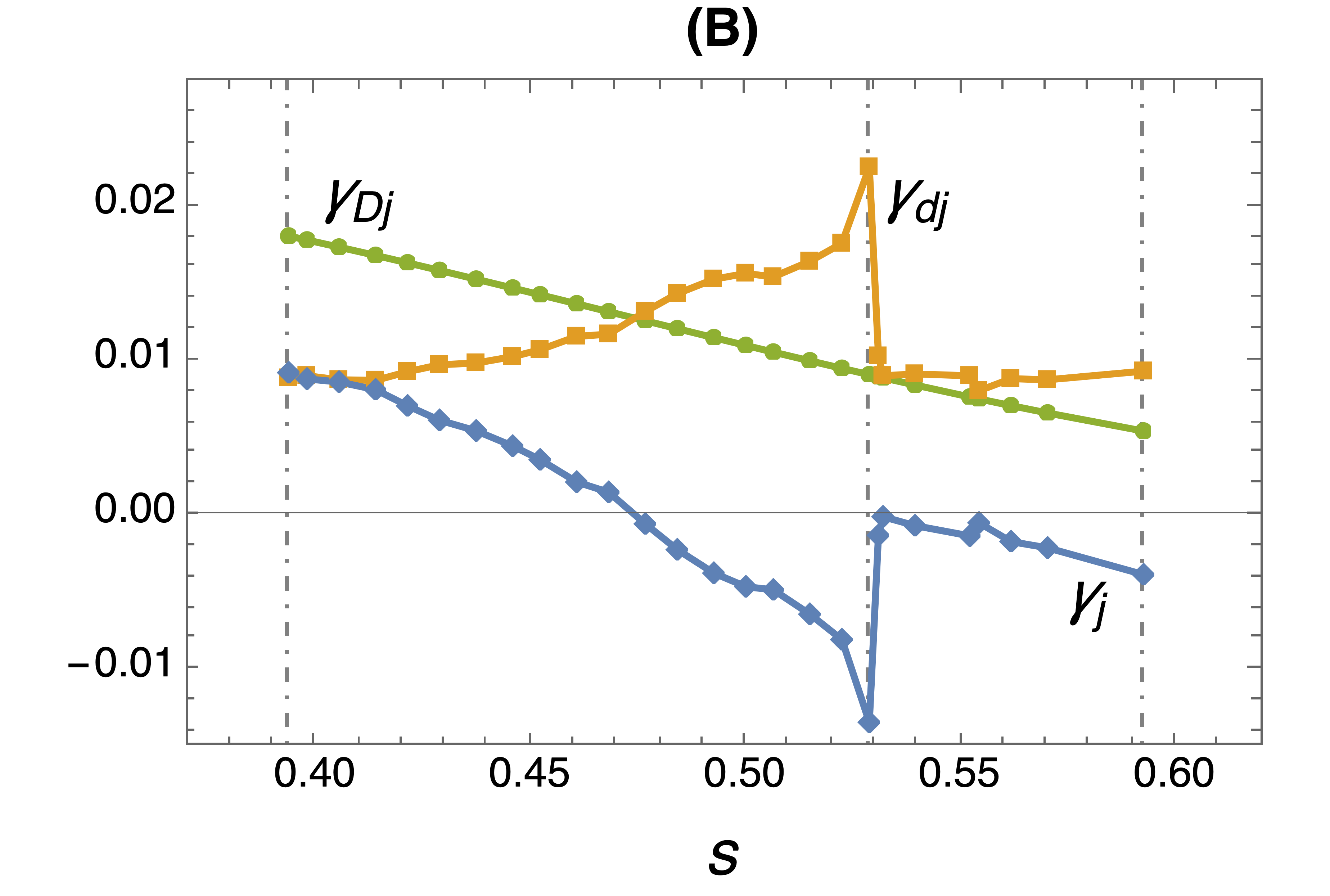}
\caption{Panel A --- Initial EP profile $f_{H}^0$ (dashed black) in arbitrary units as a function of $s$ from \eref{fhot}. Colored vertical lines indicate resonance positions: high-$n_{AE}$ branch (HnB) in red, low-$n_{AE}$ branch (LnB) in blue (for the sake of clarity, we also indicate the position of $n_{AE}=30$ and $n_{AE}=5$, the extremes of the spectrum). \;\; Panel B --- Growth rates $\gamma_j$, linear drives $\gamma_{Dj}$ and damping $\gamma_{dj}$ (scaled data from \cite{spb16}), as a function of the resonance position of the corresponding mode. $\gamma_{Dj}$ are evaluated from \eref{disrel} with $\eta=0.007$. Gray dashed vertical lines indicate the HnB and LnB resonant regions.  }
\label{fig_1}
\end{figure}

The ITER 15MA baseline scenario is taken as reference case \citep{spb16} (see also \cite{ML12,ML13,Ph15}). The initial EP profile is described by \eref{fhot} with $B_1\simeq0.5$, $B_2\simeq-1.2$ and $B_3\simeq3.2$, and the least damped 27 toroidal AEs, described in the linear analysis of \cite{Ph15}, are included in the dynamics. Specifically, $n_{AE}\in[12,30]$ for the main (high-$n_{AE}$) branch (HnB) and $n_{AE}\in[5,12]$ for the low branch (LnB). In \figref{fig_1}---A, we plot $f_{H}^0(s)$ from \eref{fhot} indicating with colored vertical lines the mode resonance positions. The determination of the free parameter $\eta$ is outlined in some detail in \cite{carlearo-ppcf,nceps22} and it results related to the fraction of the most resonant particles. Consistently, we assume $\eta=0.007$ (with $\ell^*=400$) throughout this paper. In \figref{fig_1}---B, we show the obtained linear drives from \eref{disrel} using this value for $\eta$ and the above mentioned $f_{H}^0(s)$. We also plot the damping $\gamma_{dj}$ obtained by scaling data from \cite{spb16} (as previously described), and the corresponding effective growth rates $\gamma_{j}$. It clearly emerges how part of the HnB and all the LnB results linearly stable (sub-dominant). The damping, growth rates, and frequencies of the considered modes are thoroughly examined in \cite{Ph15}, which highlights the complexity of the damping behavior. In particular, the least damped TAE branch changes from $n_{AE}=m,m+1$ to $n_{AE}=m+2,m+3$ (here $m$ denotes the poloidal harmonic mode number) with increasing  radial position ($s=0.53$), avoiding significant continuum damping for $n_{AE}<12$.

\begin{figure}
\centering
\includegraphics[width=0.475\textwidth]{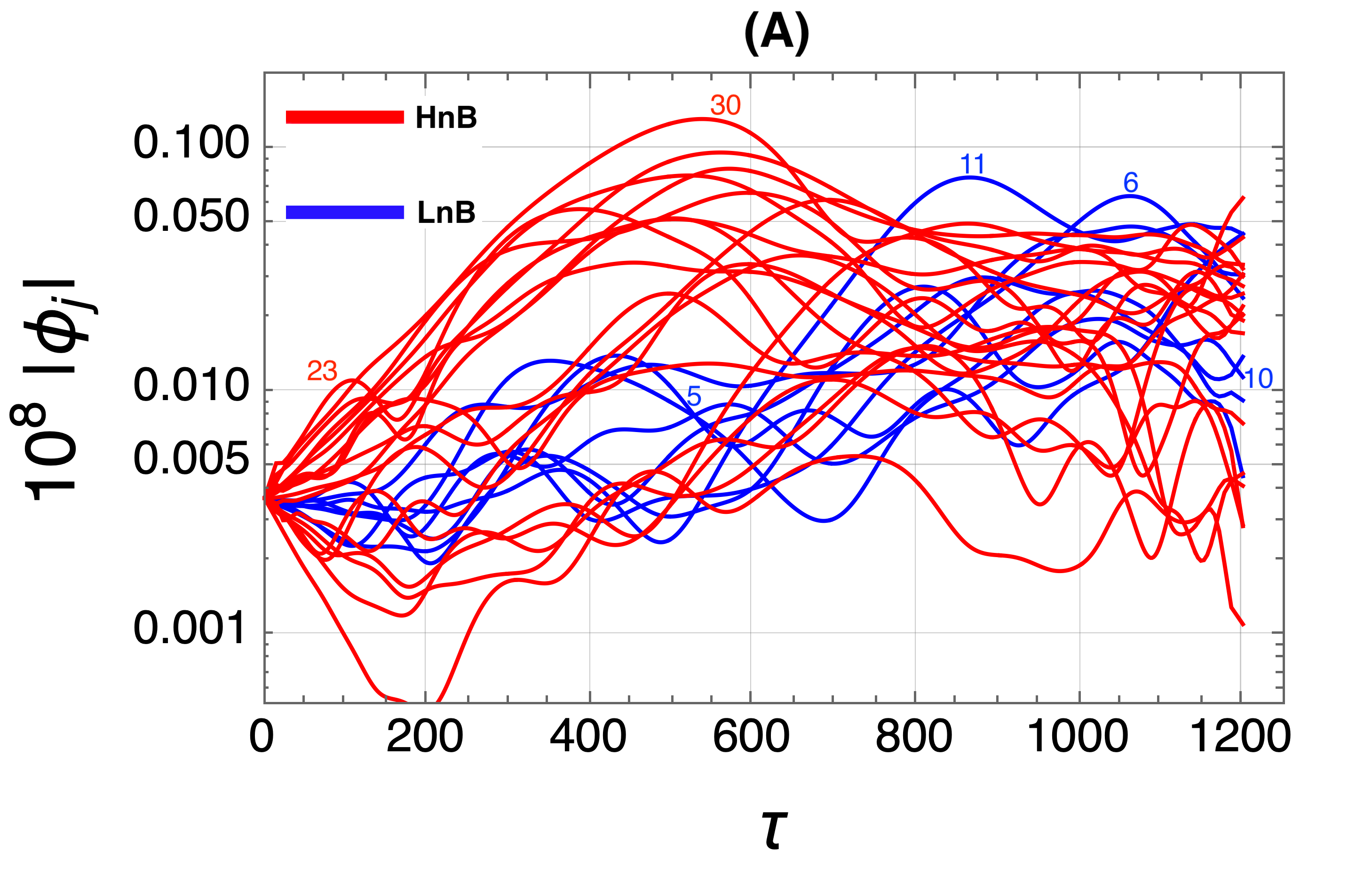}\\
\includegraphics[width=0.45\textwidth]{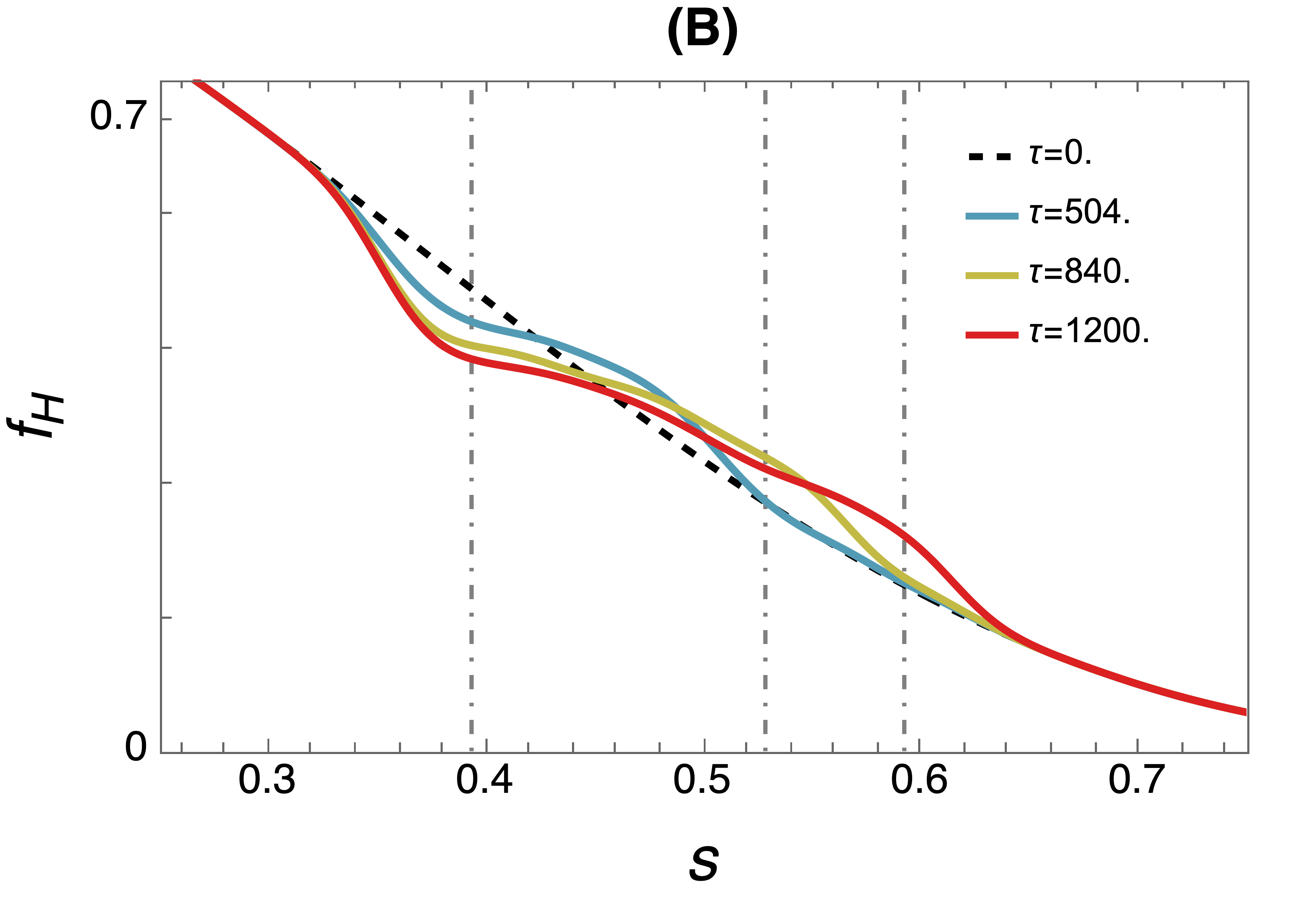}\;\;
\includegraphics[width=0.45\textwidth]{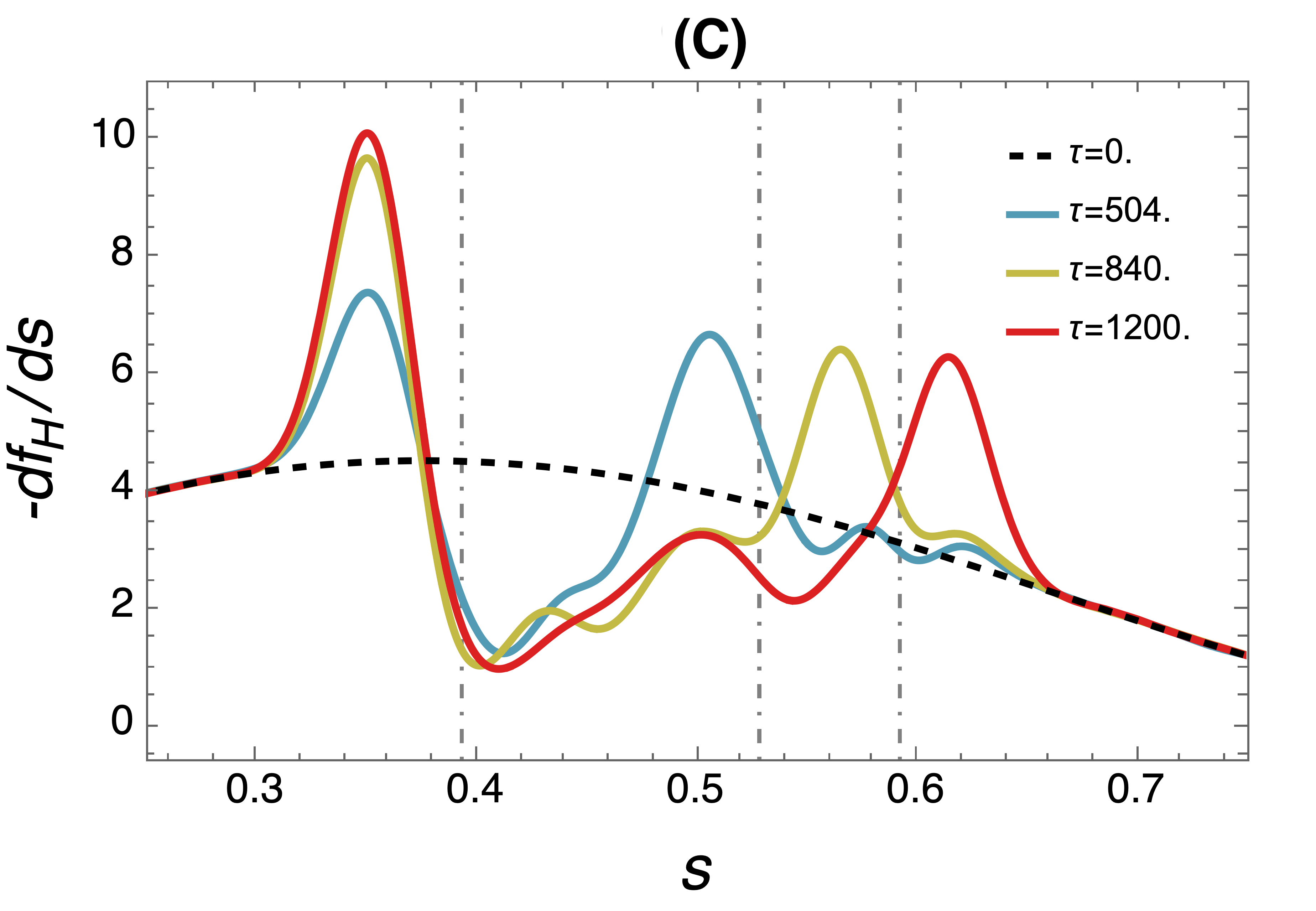}
\caption{Panel A --- Self-consistent evolution of the 27 electrostatic modes $|\phi_j(\tau)|$, which correspond to the reduced representation of the AEs (examples of mode numbers are also indicated).\;\; Panel B --- Evolution of EP profile as function of $s$ (different times are indicated in the plot). Gray dashed vertical lines indicate the HnB and LnB resonant regions. Panel C --- Evolution of the radial gradient of the EP profile \cite{carlearo-ppcf}.}
\label{fig_2}
\end{figure}

Using this setup, \erefs{mainsys1} are integrated in time\footnote{In the nonlinear simulations of this work, we use a Runge-Kutta (fourth order) algorithm and we set $N=10^{6}$, while $x_i(\tau=0)$ are given uniformly between $0$ and $2\pi$ and modes are initialized at $\mathcal{O}(10^{-14})$, in order the initial linear regime is guaranteed.} and the resulting temporal mode evolution is plotted in \figref{fig_2}---A. Notably, the stable part of the spectrum is now destabilized resulting excited by a domino mechanism (in agreement with \cite{spb16}). This kind of excitation is evident when observing the evolution of the distribution function (\figref{fig_2}---B) and its derivative with respect to $s$ (\figref{fig_2}---C). We recall that the derivative with respect to the radial quantity plays here the role of the well-know velocity derivative of the BoT model, thus representing part of the drive for the mode instability. In particular, a first peak emerges around $s\simeq0.35$ (related to the excitation of part of HnB), while a second peak (starting in the region of the sub-dominant LnB) is shifting in time toward large $s$ values. This feature characterizes the avalanche transport toward the plasma edge underlining the relevance of the sub-dominant part of the spectrum and, in this work, we provide a detailed description of such a mechanism. In \citep{carlearo-ppcf}, it is also found that the QL model fails in describing this scenario: when the QL approximation is implemented, the resulting distribution function plateau is much narrower compared to the nonlinear simulations and localized to the center of the spectrum. This is due to the fact that the sub-dominant modes are not destabilized. However, it is important to emphasize that, in general, QL model can describe the excitation of modes close to marginal stability, thus generating a broadening of the plateau. This is due to the steepening of distribution function at the edge of the plateau which leads to an increase of the drive. Nonetheless, in our specific scenario, the magnitude the damping rates brings to a peculiar situation where the QL predictability is reduced \cite{carlearo-ppcf}. The observed features can be attributed to the strong damping values depicted in \figref{fig_1}---B: the QL destabilization mechanism alone (pure diffusive dynamics) has not the capability to excite the sub-dominant portion of the spectrum, and this leads to the discrepancy described above.

\subsection{Analytically constructed benchmark case}\label{rd-case}
\begin{figure}
\centering
\includegraphics[width=0.47\textwidth]{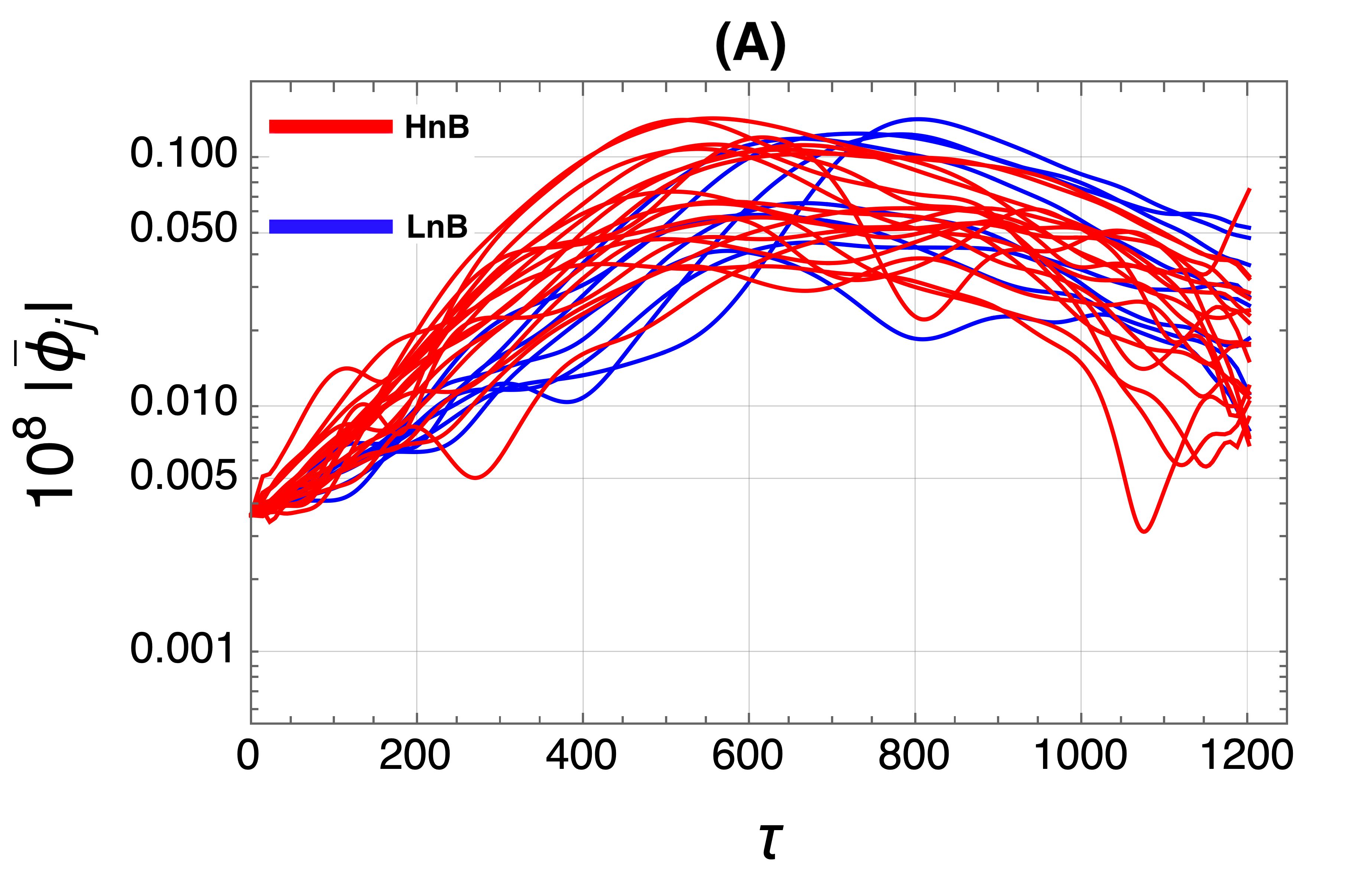}
\includegraphics[width=0.435\textwidth]{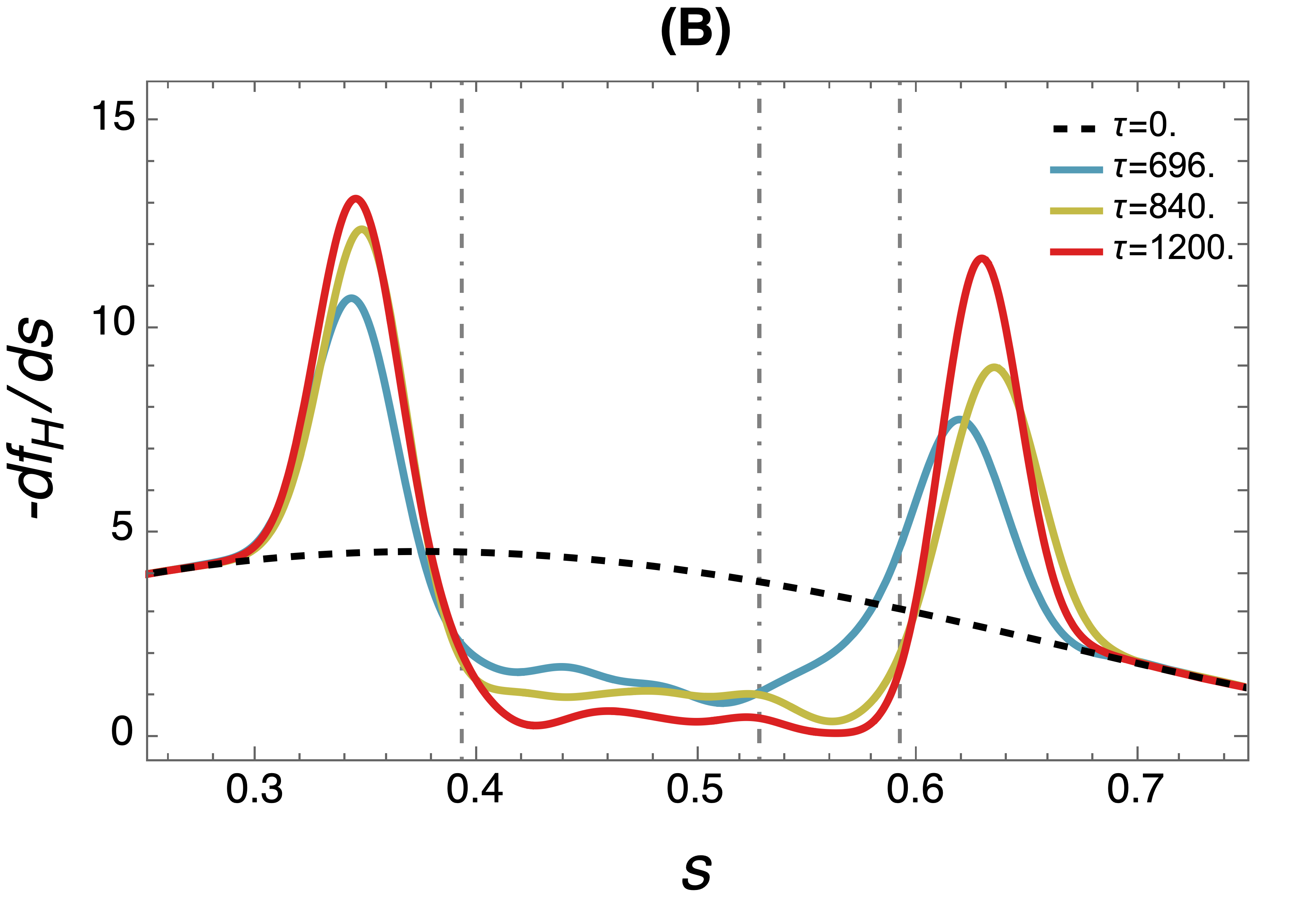}
\caption{Benchmark case. Panel A --- Plot of the mode self-consistent evolution.\;\; Panel B --- Plots of the radial gradient of the EP profile. Gray dashed vertical lines indicate the HnB and LnB resonant regions.}
\label{fig_4}
\end{figure}
We conclude this section by introducing a second case that serves as a comparative study, derived from the ITER 15MA baseline scenario, in which the stable part of the spectrum is made artificially unstable. Specifically, we assume \textit{ad hoc} reduced damping values but maintaining equal drives: since the effective growth rates are $\gamma_j=\gamma_{Dj}-\gamma_{dj}$, in this case all modes result linearly unstable, also the LnB. This allows a comparison with the realistic ITER scenario, in fact all modes are now simultaneously excited resulting in the evolution outlined in \figref{fig_4}---A, where the domino effect is negligible. We stress how the mode saturation levels, by virtue of the growth rate profile, are almost identical to the previous scenario. The evolution of the distribution function gradient is shown in \figref{fig_4}---B: the shifting peak which was related to the avalanche dynamics in the ITER scenario is now consistently localized around $s=0.64$.

\section{Phase-space transport}\label{sec4}
When the transport is primary driven by resonance effects between particles and modes, the resulting structures characterizing particle dynamics in phase space due to the potential resonant trapping (such as invariant manifolds) play a crucial role and several techniques can be used for their description. However, this task is inherently complex, as the fluctuation amplitudes have a non-trivial time dependence that must be calculated self-consistently with the evolution of the distribution function. Therefore, great care must be used when applying standard techniques typically used for analyzing periodic dynamical systems, such as surface of section analysis, that require a generalization. The concept of LCS \citep{Haller_2015} has been introduced to characterize transport processes in complex fluid flows with similar features. LCS generalize structures observed in autonomous and periodic systems, e.g. stable and unstable manifolds, to temporally non-periodic flows. Analogously to these structures, LCS divide the phase space into macro-regions inside where fast mixing phenomena take place.

As described in the introductory section, following \cite{jpp_cmf,CFMZJPP}, we will use the LCS technique to characterize the dynamics in the phase space illuminating the relaxation process described in the previous section. Specifically, we will focus on hyperbolic LCS that organize the flow by either attracting or repelling volume elements over a finite time span, generalizing the concept of stable and unstable manifolds. Various methodologies have been proposed to compute hyperbolic LCS, see e.g. \cite{H11} or \cite{SLM05}. Since the largest Finite-Time Lyapunov Exponent (FTLE) at a specific phase space location quantifies the exponential separation between two neighboring initial conditions after a certain time interval $\Delta \tau$, hyperbolic LCS can be qualitatively described extracting 1D structures with peaked FTLE profile, see e.g. \cite{MPJPP}. However, how Haller has shown in \cite{H11,H12}, looking at these ``ridges'' of the FTLE profile may sometimes fail in finding the exact shape of the LCS thus requiring a more robust definition, see e.g. \cite{Pegoraro_2019} for a detailed discussion. Nonetheless, in this work, we plot the contours of the FTLE field to extract an immediate information about the dynamical feature of our system and, in particular, distinguish between convective/ballistic and diffusive dynamics. We then evaluate their robustness by over-plotting the tracers evolution alongside the barriers.

\begin{figure}
\centering
\includegraphics[width=0.7\textwidth]{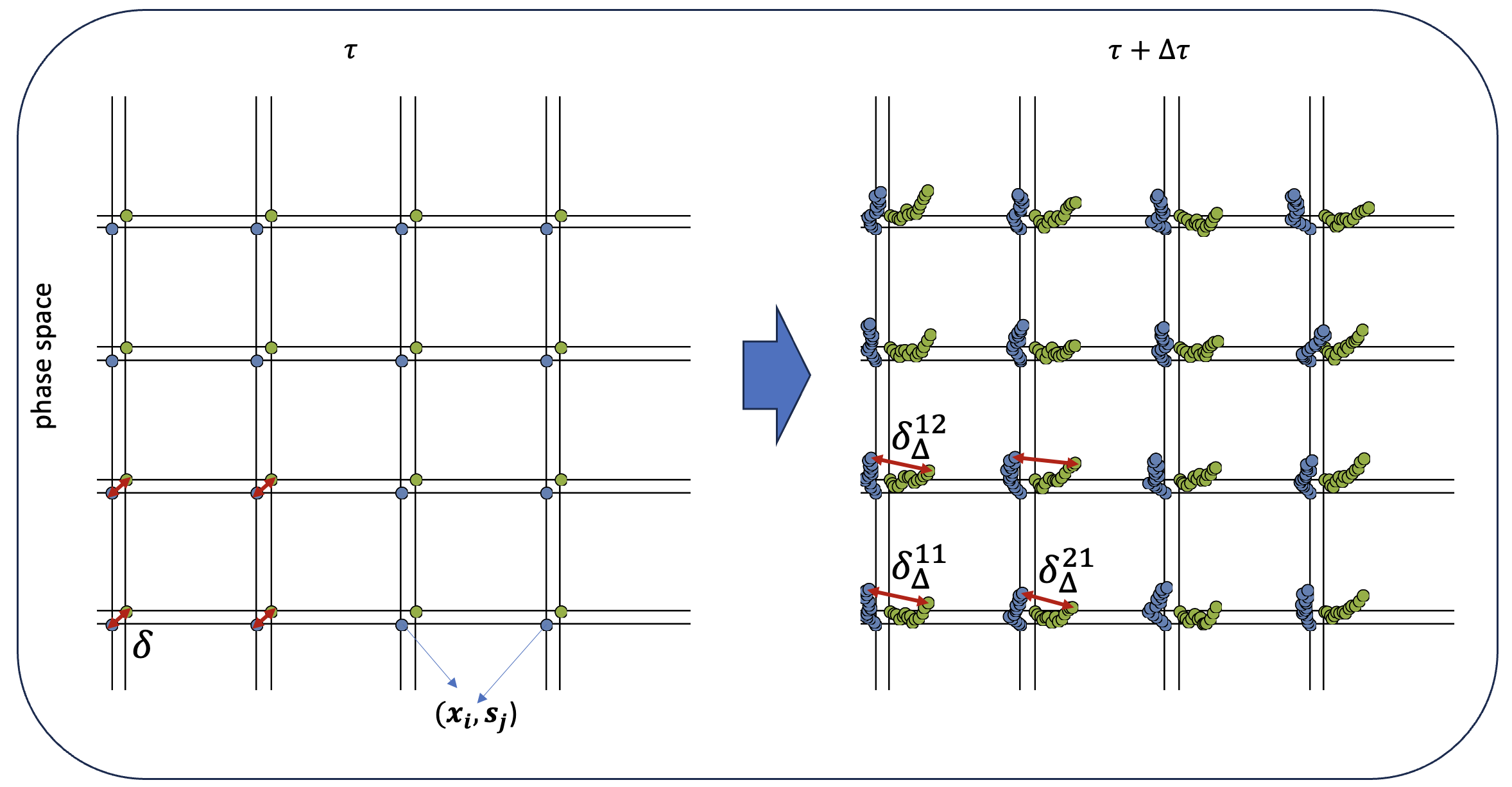}
\caption{Cartoon representing the Finite Time Lyapunov Exponent calculation consistently with Eq.(\ref{ftle}).}
\label{fig_10}
\end{figure}

In the following, we employ a test particle approach and we stress that we use, accordingly to the previous section, the notation for the phase space as ($x$, $s$) to well underline the EP redistribution in comparison with respect to HAGIS-LIGKA simulations. In fact, the variable $s$ is treated in the mapping scheme as a velocity variable, and this define without ambiguity the reduced phase space we are considering. Tracer trajectories are evaluated using the assigned potential fields described in the previous sections, i.e. tracers are evolved by means of Eqs.(\ref{eq1}) and (\ref{eq2}) alone with time dependent potential extracted from the fully non linear simulations. We track markers initialized on two phase-space grids at a given time $\tau$, as depicted in \figref{fig_10}. Points of the first grid are denoted as $(x_i,s_j)$ while points on the second grid as $(x'_i,s'_j)$ obtained from the first one through a small translation $\delta$. Each grid is 400$\times$400 for a total of 320000 evaluated tracers. This choice has followed a resolution scan in order to obtain well peaked structures. Moreover, since $x$ variable is $2\pi$ periodic, the grid are initialized covering roughly $2\lambda_{max}$, where $\lambda_{max}=2\pi/\ell_{min}$ (with $\ell_{min}=\text{Min}[\ell_j]$). As the tracers evolve, by the time $\tau+\Delta\tau$, the distance between a point and its corresponding point on the second grid can be denoted as $\delta_{\Delta}^{ij}$. The FTLE field $\sigma_{i,j}(\tau,\Delta \tau)$ is readily computed on the first grid as
\begin{equation}
\label{ftle}
\sigma_{i,j}(\tau,\Delta \tau)=\ln(\delta_\Delta^{ij}/ \delta)/\Delta \tau.
\end{equation}
As previously mentioned, hyperbolic LCS are closely linked to the contours of $\sigma_{i,j}$ in the phase-space, especially with 1D peaked structures known as ridges \citep{SLM05}. It is well known that, in a scenario involving a single mode with a constant amplitude, similarly to a pendulum, attractive and repulsive structures will merge into the separatrix. The hyperbolic LCS will reflect this, as shown in \cite{jpp_cmf, CFMZJPP}. Considering a general non-autonomous dynamical system, multiple intersections occur allowing transport by means of the so called lobe-dynamics, see e.g. \cite{Di_Giannatale_2018}.

We calculate FTLE contours for both the ITER scenario and the benchmark case. For a given $\Delta\tau>0$, FTLE peaks indicate repulsive barriers while attractive barriers are obtained with the same methodology by considering the backward time dynamics. This can be associated to negative values of $\Delta\tau$. We plot the contours of the FTLE using $\Delta\tau=\pm200$ and mode amplitudes at different times. In order to show the results of both contours on the same plot, for each point of the phase space, the maximum between the two FTLEs field is plotted. This method is effective, as demonstrated in \cite{CFMZJPP,jpp_cmf}. The accuracy of the obtained structures is then verified by following a set of passive tracers at different times. The choice of $\Delta \tau$ requires care, larger $\Delta \tau$  generate contour plots with long structures approximating the well-known tangles between stable and unstable manifolds. Smaller values result in simpler structures describing the coherence of the mixing of passive tracers over a shorter time interval, see e.g. \citep{Di_Giannatale_2018}, for a detailed discussion regarding this point. After several sensitivity tests, the selected value of $\Delta \tau$ produces FTLE fields exhibiting sufficiently pronounced peaks to clearly identify the LCS; moreover, the superposition of passive tracers provided an additional benchmark for the effective transport barriers, supporting the choice of $\Delta \tau$. In the following, we illustrate how these structures organize the dynamics in the phase-space by analyzing the evolution of markers initialized at  $n_{AE}=5,\,12,\,30$ resonances.

\begin{figure}
\centering
\includegraphics[width=0.95\textwidth]{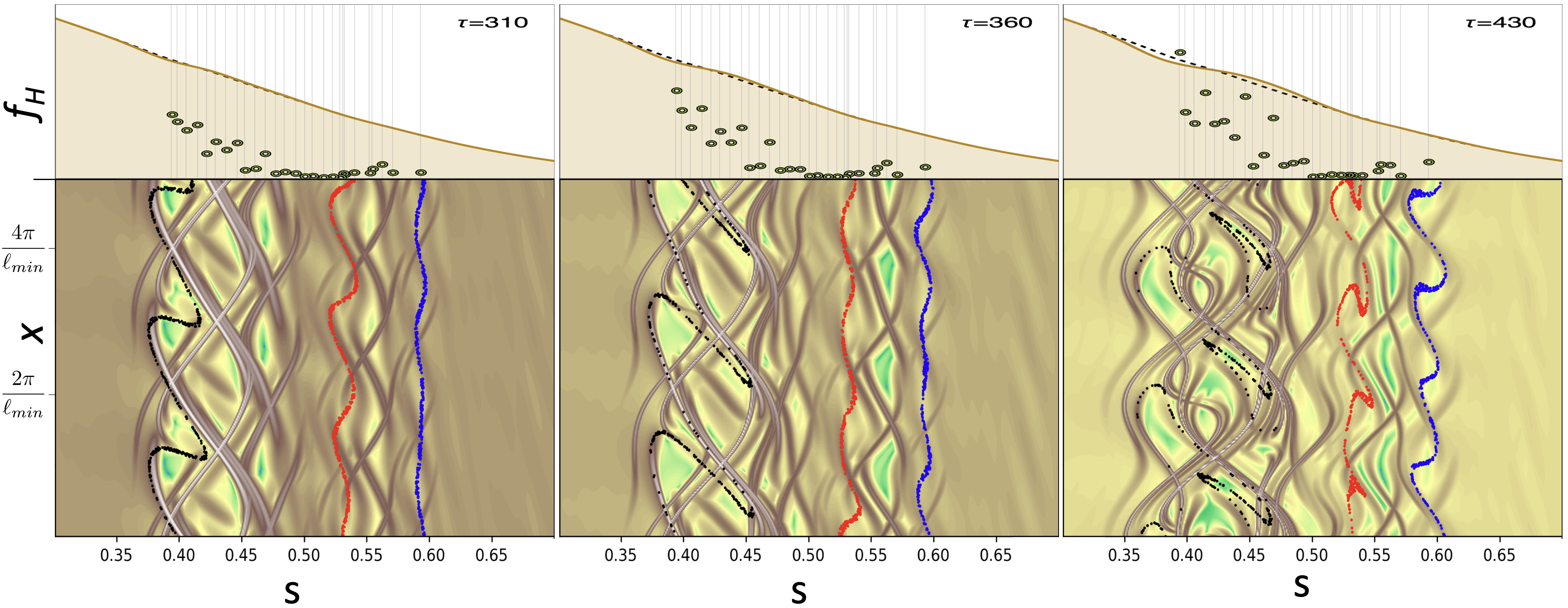}
\caption{ITER scenario. Contours of the FTLE in the initial phase of the dynamics are plotted in the lower panels with the following color-scale: \includegraphics[draft=false,width=5cm]{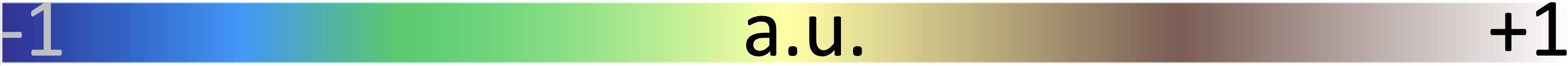}. The evolution of the distribution function (we also represent the initial profile with a dashed black line) overplotted on the mode amplitudes (green annulus) is depicted in the upper panels in arbitrary units. The dynamics of three marker populations initialized around the $n_{AE}=5$ ($s\simeq0.59$, blue), $12$ ($s\simeq0.53$, red) and $30$ ($s\simeq0.39$, black) are overlplotted. The periodicity length in the $x$ variable is $2\pi$ and, to well highlight the FTLE structures, the plot scale is limited roughly $4\pi/\ell_{min}$ (with $\ell_{min}=\text{Min}[\ell_j]\simeq660$).}
\label{fig_11_early}
\end{figure}

\begin{figure}
\centering
\includegraphics[width=0.95\textwidth]{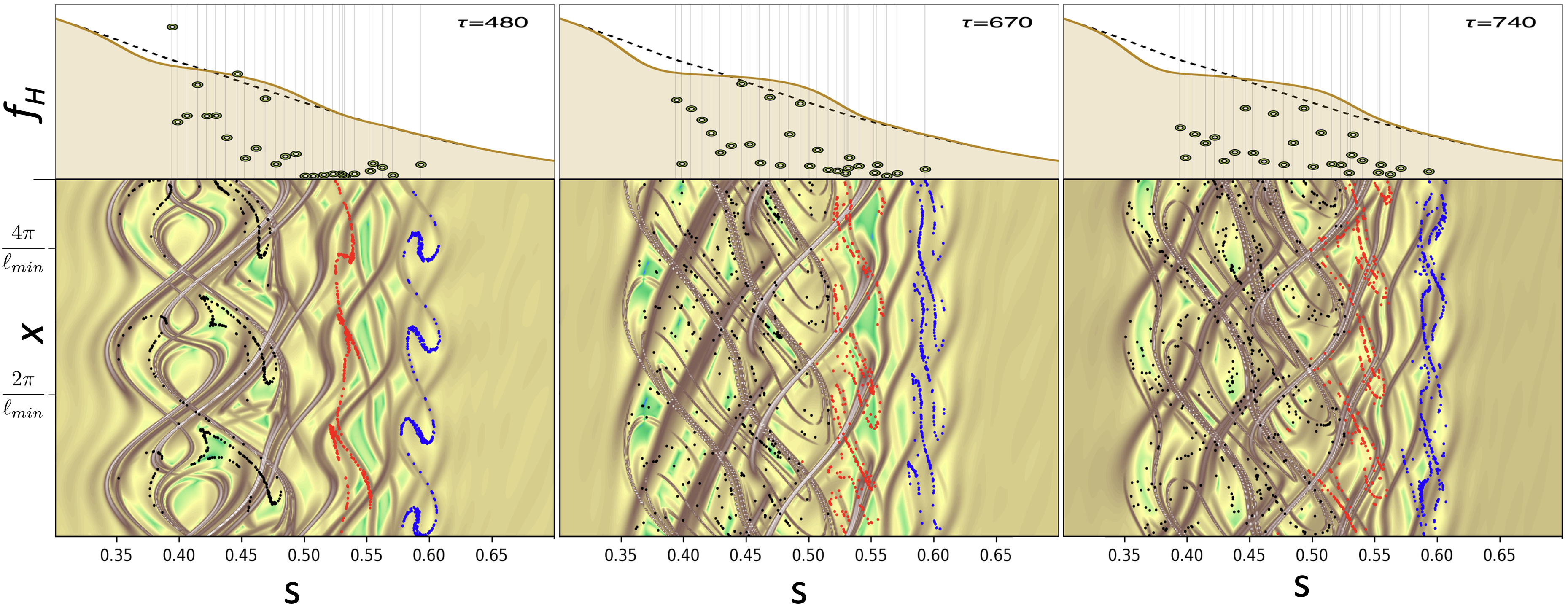}
\caption{ITER scenario. Same as \figref{fig_11_early}, but for the late phase of the dynamics.}
\label{fig_11_late}
\end{figure}

The analysis of ITER scenario is summarized in \figref{fig_11_early} and \figref{fig_11_late}. We emphasize that we show the evolution of the distribution function $f_H(\tau,s)$ evaluated as the continuous histograms in $s$ (averaged over the dimension $x$). This corresponds to proper construction of the distribution function in our $N$-body Hamiltonian representation we are using in this work. At $\tau = 310$, passive tracers rotate within single resonance structures, as expected \citep{TMM94,jpp_cmf}. Such a rotation of the passive tracers is representative of the real EP motion: when EP get resonantly trapped the energy is transferred to the modes increasing their amplitude. This process drives an elongation of attractive and repulsive structures that cover $\sim 10$ resonances at $\tau = 360$ and particles move almost coherently along these structures. Before $\tau \simeq 430$, the FTLE contour plots do not exhibit strong ridges in the region $0.49\lesssim s \lesssim 0.53$, where we recall that $s=0.53$ is the separation between HnB and LnB. This region is characterized by strongly damped modes of the HnB and their subsequent destabilization corresponds to the start of the avalanche process. We underline the fact that, instead, the lower ridges found for $0.53\lesssim s\lesssim 0.60$ are due to a nonlinear destabilization (at very small saturation levels) of the linearly stable LnB as can be noted by observing the spectrum. Mixing phenomena within the plateau region increase the absolute value of the distribution function gradient and consistently the drive of the HnB modes with $s\simeq 0.50$, along with the elongated ridges of the FTLE field.

By $\tau = 430$ attractive and repulsive structures intertwine into a single tangle with $0.35\lesssim s \lesssim0.49$. We recall that the presented snapshots are evaluated at a single time frame but they include the dynamics of tracers in the time interval $\tau\pm200$. Thus, the observed intertwining and the related ability of particles to interact with adjacent modes constitutes the base of a diffusive behavior in this region (which corresponds to the HnB resonant portion) that should occur around $\tau=600$. Consequently, the distribution function will slowly form a plateau. In fact, by $\tau = 480$, FTLE ridges cover a significant portion of the phase-space initially characterized by stable modes. Consistently, these modes are driven, and the plateau extend to the right while the HnB transfer energy to the LnB. Moreover, tracers resonating with the HnB (black and red) cover a large region, thus interacting simultaneously with multiple resonances. The trend continues at $\tau = 670$ and $\tau =740$, leading to an expansion of the region where fast mixing phenomena take place. Once all the modes have sufficiently large amplitude, the phase space is characterized by a unique plateau.

As previously stated, the LCS technique enables the identification of dynamically evolving transport structures in a fully nonlinear and time-dependent environment, thereby providing a powerful diagnostic tool for characterizing the intrinsically intermittent and avalanche-like nature of transport processes in magnetically confined fusion plasmas. In our analysis, we identified and characterized the phase-space structures underlying the distribution relaxation process. These structures progressively deform and stretch in time, increasing their overlap across extended regions of phase space. As a consequence, confinement is locally weakened and particle redistribution is enhanced. From this perspective, the observed transport can be understood as the result of the dynamical evolution and progressive elongation of phase-space structures, which ultimately regulate the onset and development of avalanche-like relaxation events.

Nonetheless the excitation phenomenon of linearly stable modes can be addressed in the standard QL theory, the high values of damping considered in the reference scenario reduces its predictivity for this specific case, as already described. Specifically, the generated stiffness of the distribution function gradient is not sufficient to generate positive effective growth rates during the evolving dynamics. In this respect, we note how modified QL pictures can describe scenarios strongly dominated by avalanche phenomena, and in future works, we aim at analyzing in detail the efficiency of such schemes. In particular, in \cite{spb16}, it is argued that the avalanche EP transport processes observed for this setup can be included in the scheme of the so-called Line Broadened QL (LBQL) model \citep{BB95b,GBG14} and its earlier Resonance Broadened QL (RBQ) extension \citep{gore18} (see also \cite{2023PhRvL.130j5101D}).

\begin{figure}
\centering
\includegraphics[width=0.95\textwidth]{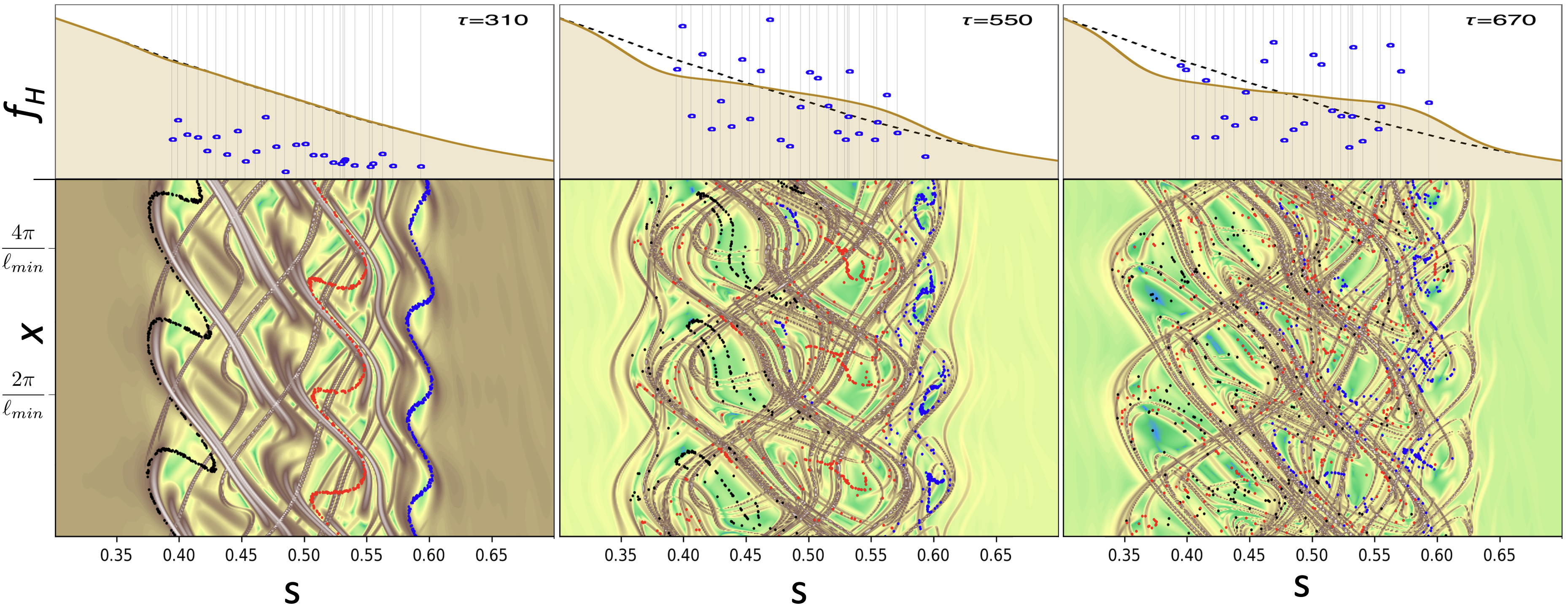}
\caption{Benchmark case. Same as \figref{fig_11_early} (here the mode amplitudes are denoted with blue annulus).}
\label{fig_12_}
\end{figure}

The same methodology can be applied to the benchmark case, and the results are shown in \figref{fig_12_}. The dynamics until $\tau \sim 550$ includes linear mode saturation and it shows how the tracers follow attractive barriers during their evolution. LCSs overlap already at $\tau = 550$: as it can argued from the figure, the peaked profiles of FTLE form structures that cover the whole resonant region. Consequently, fast mixing phenomena take place and particles start to experience the drive of different modes. The distribution function will exhibit, in turn, a unique plateau. In this sense, modes result almost simultaneously excited and particles interact with the whole spectrum during a time-scale comparable to the single mode trapping time. This corresponds to the basic underlying features of a diffusive regime of transport.

The obtained results for the analyzed ITER scenario will be further elaborated in the next section, with a more in-depth discussion of the excitation of the sub-dominant part of the spectrum, also including a comparison between statistical and kinetic frameworks.

\section{Definition of transport regimes}\label{sec3}
When addressing the ITER 15MA baseline scenario, the proposed reduced model introduced in the previous sections effectively highlights the failure of the QL model in characterizing the relaxation process \citep{carlearo-ppcf}. Anyway, we remark that the breakdown of the QL assumptions actually does not guarantee the non-applicability of the model, also when strong mode-mode coupling is present (see the review \cite{ee18} and \cite{Besse11,ql20nc,gore18}). The observed discrepancy for the ITER case under investigation, thus, requires a careful analysis of the stochasticization processes underling the distribution function evolution.
\begin{figure}[ht!]
\centering
\includegraphics[width=0.56\textwidth]{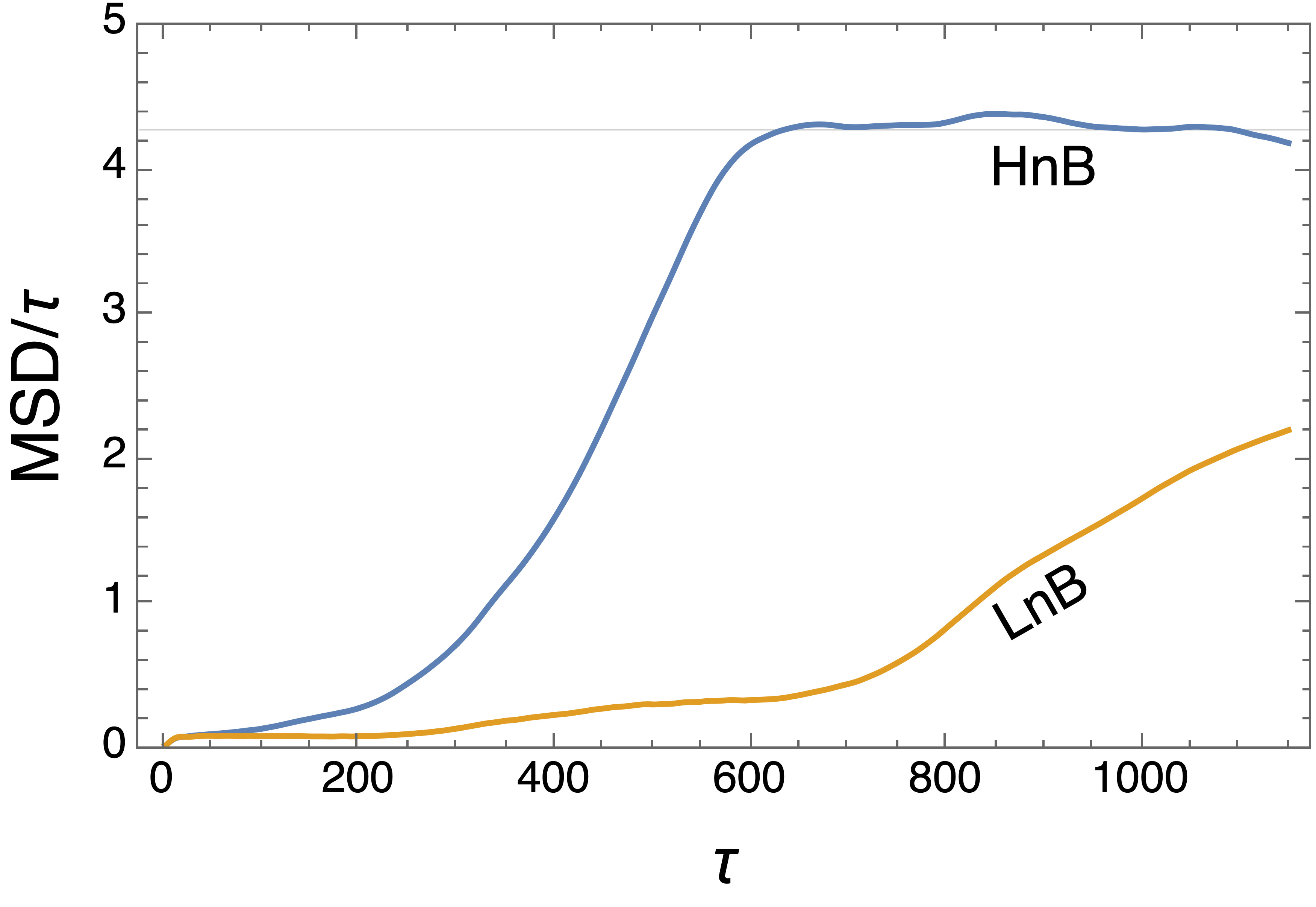}
\caption{Plot of MSD$/\tau$ (arbitrary units) as a function of $\tau$ for tracers initialized in the resonant region of the HnB (blue) and the LnB (yellow).}
\label{fig_msd}
\end{figure}

The relaxation process can again be analyzed with test particle analysis. Specifically, we track the mean square displacement as a function of time: MSD$=\langle[s(\tau)- s(0)]^2\rangle$, where averages $\langle...\rangle$ are taken over the population of test-particles. For completeness, we specify that the following analysis accounts for $10^6$ tracers for each run. Different behaviors of the MSD discriminate distinct transport features. In particular, diffusion is characterized by constant MSD$/\tau$ (for a different but complementary statistical approach, see \cite{VK12,Zagorodny03,KVG19,VK16}).
\begin{figure}[h!]
\centering
\includegraphics[width=0.6\textwidth]{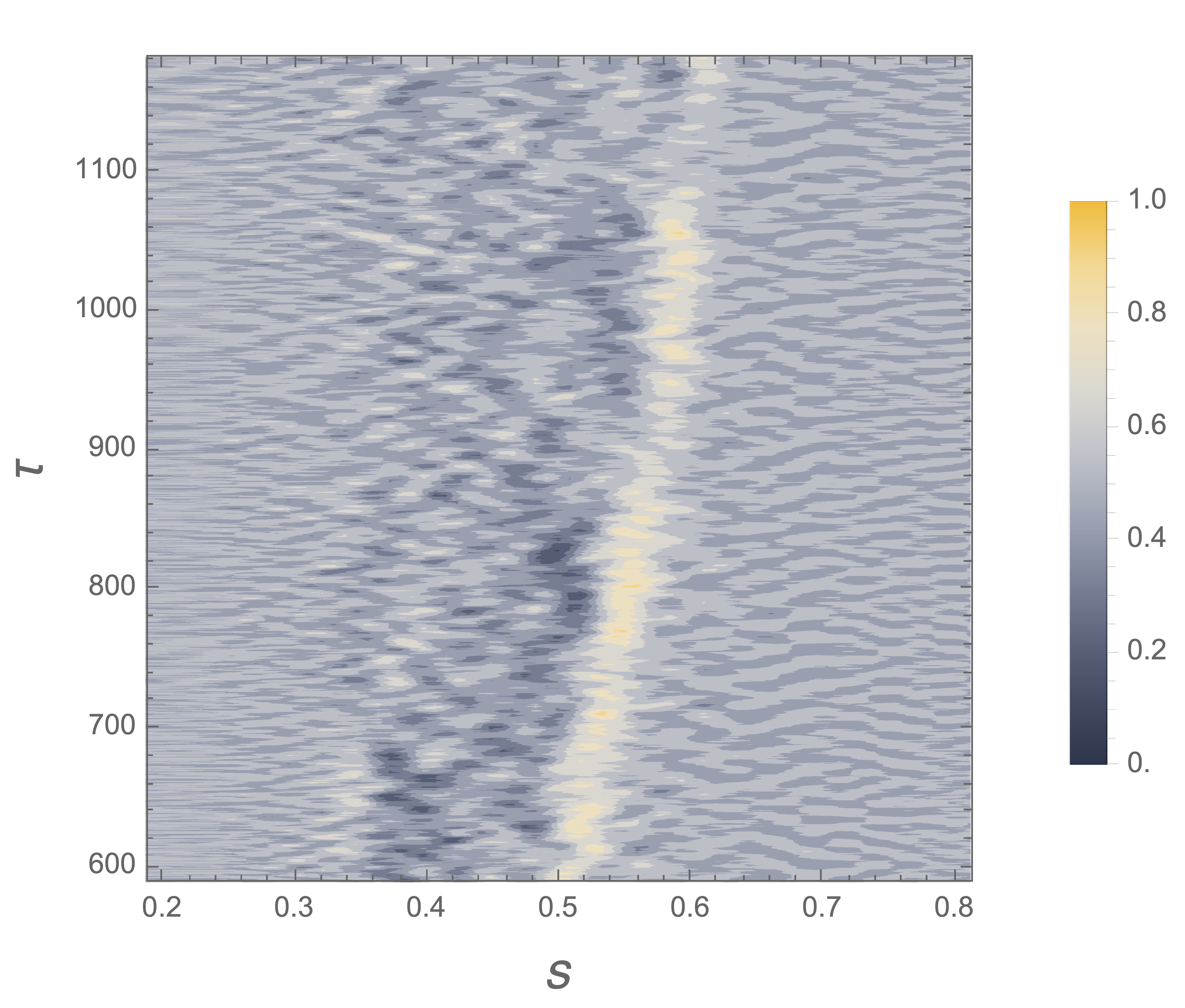}
\caption{Contours of the flux divergence (i.e. $\partial_\tau f_H(s)$) for the ITER scenario.}
\label{fig_fluxes}
\end{figure}
\begin{figure}[ht!]
\centering
\includegraphics[width=0.95\textwidth]{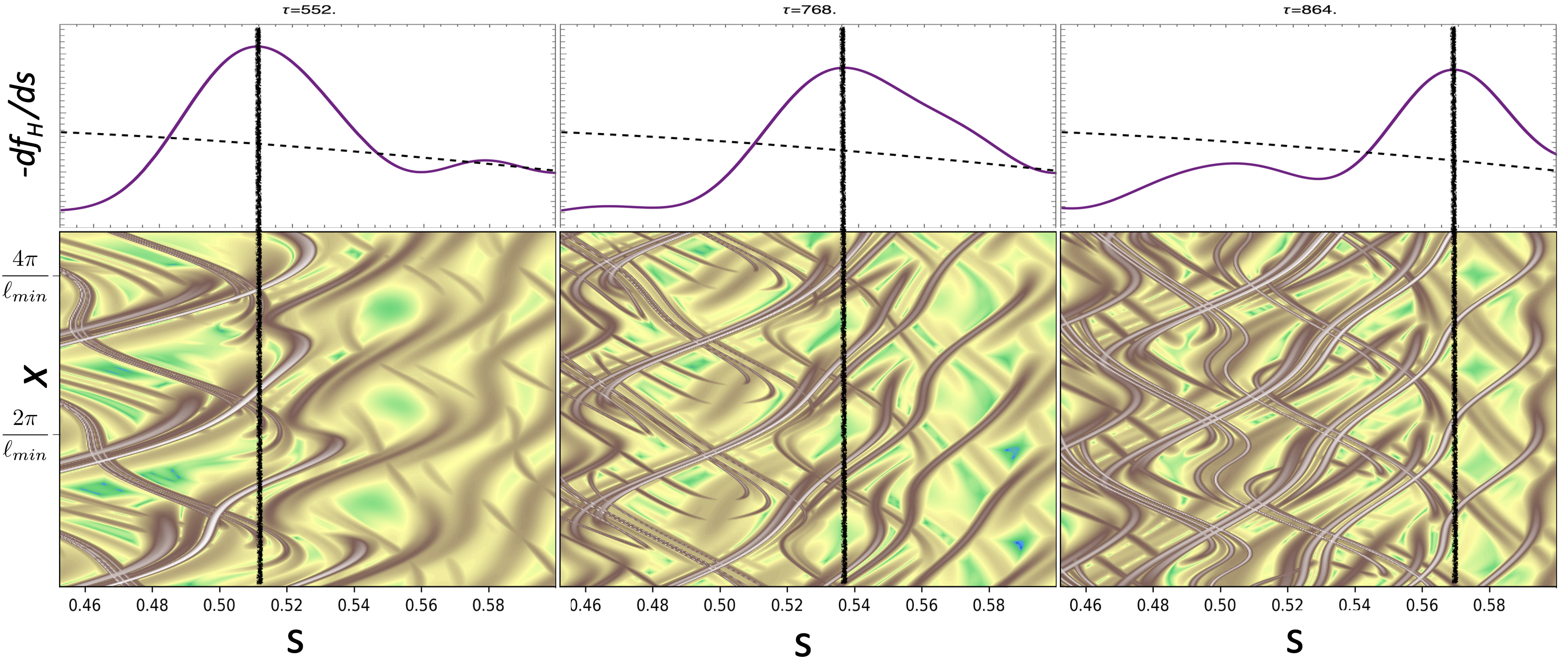}
\caption{ITER scenario. Contour plots of the FTLE as in \figref{fig_11_early} are plotted in the lower panels. The evolution of the derivative of the distribution function is reported in the upper panels (the black dashed line represents the initial time $\tau=0$). The vertical line highlights the maximum gradient position.}
\label{fig_zoom}
\end{figure}

We track the MSD for two distinct tracer populations, namely: test particles initialized in the resonant region of the HnB (from $s_{30}\simeq0.39$ to $s_{12}\simeq0.53$), and of the LnB (from $s_{12}$ to $s_{5}\simeq0.59$), thus extending the local tracer analysis in \citep{carlearo-ppcf}. This provides information on transport of specific resonant populations reflecting the underlying spectral features. In \figref{fig_msd}, we show the evolution of MSD$/\tau$ for such particles consistently with the results of the previous section. It emerges how the first population is characterized by diffusive transport during the post saturation dynamics, where MSD$/\tau$ reaches a constant value. Conversely, for the second population, for $\tau>800$ most of the sub-dominant part of the spectrum is saturated while MSD$/\tau$ continues to growth almost linearly. This indicates a non-diffusive, and in this case convective, transport regime. It is also worth stressing that, regarding the saturation time, i.e. $\tau>600$ for the HnB and $\tau>800$ (see \figref{fig_2}---A), we refer to the stage in which the majority of the modes are saturated while a negligible part of the spectrum persists in a nonlinear regime.

In order to provide further evidence of non diffusive features, let us now analyze the particle fluxes for the ITER scenario. The flux divergence, i.e.  $\partial_\tau f_H(s)$, is evaluated by finite differences and is represented in \figref{fig_fluxes}. Transport within the plateau region, where resonances are strongly overlapped, can be reasonably described within a diffusive framework. In contrast, the shoulder of the distribution function appears to drift coherently toward higher $s$ values with approximately constant velocity, as clearly seen in \figref{fig_fluxes} (and also in \figref{fig_zoom}). The LCS analysis indicates that this behavior is due to an attractive structure that moves coherently toward larger $s$ as a consequence of the progressive destabilization of the spectrum. Furthermore, the elongated nature of this attractive structure is associated with a streaming motion of particles along the structure. Both effects are signatures of non-diffusive transport. This interpretation is consistent with the MSD analysis, which clearly shows non-diffusive behavior for the corresponding particle population. 

In \figref{fig_zoom}, we plot the FTLE contours together with the gradient of the distribution function. We first note that the snapshots presented here are obtained by monitoring the tracers over the interval $\tau\pm200$, thus including the mode saturation phase described above. The highest peaks of the FTLE field follow the shifting of the distribution function gradient, which results in the domino excitation mechanism. This is consistent with the fact that the drive is maximum in this region. The adoption of the LCS methodology provides a direct and physically transparent picture of the mechanism underlying avalanche-like transport. Rather than simply increasing fluctuation levels, the steep gradient in the distribution function at the boundary of the plateau preferentially drives the elongation of large-scale phase-space structures, stretching them radially. This elongation is a key element of the transport process. As the LCS extend over larger radial domains, they facilitate the redistribution of particles toward the plasma edge. In other words, transport is not merely intensified locally but becomes more organized and nonlocal in character. The stretched geometry and the enhanced peaks of the FTLE increases allows the coupling of distant radial locations, effectively acting as channels that promote avalanche-like events. At the same time, the sharp gradient feeds energy into sub-dominant spectral components through nonlinear wave–particle interactions. These interactions further reinforce the extended structures, sustaining their elongated morphology and contributing to a broader redistribution of spectral energy.

This scenario differs substantially from a constructed benchmark case in which the spectrum is linearly and globally unstable. In that case, transport arises primarily from the linear growth of unstable modes and does not rely on the self-consistent elongation and nonlinear reinforcement of coherent structures. Here, by contrast, the elongation of LCS emerges as the central dynamical feature governing enhanced and nonlocal transport.

We underline how in \cite{BB95b} avalanche phenomena are successfully described using a pure diffusive scheme (RBQ model) while this analysis indicates that the domino excitation of the sub-dominant part of the spectrum is linked to a more complex transport phenomenology. This fact could suggest that, although the QL diffusive schemes can address the domino main features, actually the fully nonlinear relaxation seems to be also accompanied by a non-zero collective ballistic transport. By other words, the present analysis stimulate the study to properly describe the transition from convective to diffusive regimes in Fokker-Planck-like schemes \citep{ncentropy}.

\section{Concluding Remarks}
In this work, we have provided a detailed examination of the transport mechanisms associated with avalanche processes emerging in the ITER 15MA baseline scenario. Drawing the contours of the FTLE together with test particles sampling the phase space, provided a clear visualization of the attractive/repulsive structures. The emergence of transport channels, driven by the formation of a steep gradient in the right tail of the plateau region, underscores the role of nonlinear particle-mode interactions. This mechanism strongly differs from the diffusive benchmark introduced, where transport is driven by a globally unstable spectrum leading to strong mode overlaps. The relaxation process has been analyzed by means of the MSD showing how particles exhibit convective/ballistic transport. The ballistic nature of particle transport is also highlighted by the analysis of the fluxes where a peaked structure moving with approximately constant velocity characterizes the dynamics. Concluding, the presented analyses identified specific factors associated with avalanche phenomena, crucial for predicting the onset of non-diffusive transport. These indicators enhance the ability to forecast distinct transport regimes without using nonlinear gyrokinetic simulations.

We remark that the model used in the present work does not include EP sources and sinks. This precludes the reconstitution of the perturbed EP distribution at the resonance and only the mechanism of distribution flattening is accounted for. We acknowledge this limitation in making saturation level predictions relevant for real experiment, but this study provides an initial application of the FTLE technique that will be relevant for more realistic applications. In future work, we plan on addressing how \textit{(i)} resonance overlap and \textit{(ii)} collisions separately contribute to making a random phase approximation valid for the kinetic system (thereby in making QL theory justifiable). The conventional QL theory requires many overlapping modes for that. In \cite{2019PhPl26l0701D}, it has been formally shown that, even for a single mode, a QL theory emerges from nonlinear kinetic theory in the case of near marginal stability (when effective collisionality is much larger than the bounce frequency). The issue that could now be addressed is what happens when there is a combination of both for which a Langevin formulation could be employed. An outstanding question to be investigated is thus how much can the constraint of many modes and/or dominant collisionality be relaxed but with the random phase still being a good approximation. Both scattering collisions and overlap produce stochastization and de-correlation of fast ions from resonances. Understanding the transition between different transport regimes, and thereby, understanding the region of validity of reduced models, can be illuminating for transport studies in both plasmas and self-gravitating systems. In the dark matter resonant dynamics, realistic cases were found to be in-between the regimes of diffusive and convective scenarios \citep{hamiltonAPJ}, which only highlights the need for future studies to analyze what happens in this poorly explored intermediate zone where neither the isolated resonance nor the fully overlapped resonances scenario occur. The analysis technique used in the present paper, employing a formulation of the collisional problem using Langevin equations, could be used. Alternatively, there is a new technique, developed to distinguish between regular and chaotic orbits, called Weighted Birkhoff Average \citep{Duignan23} that can lead to new physics insights. We conclude by stressing the importance of studying wave-wave nonlinearities \cite{todo10}, which can excite zonal structures leading to steeper EP profiles. Specific developments can be linked to reduced approaches inherent to an energy-conserving scheme such as the one presented in \cite{barberis2025}, where zonal perturbations are treated with a beat-driven generation mechanism.

\section*{Acknowledgment}
The authors would like to acknowledge insightful discussions with Prof. F. Zonca and Dr. Vin\'icius N. Duarte. This work has been carried out within the framework of the EUROfusion Consortium, funded by the European Union via the Euratom Research and Training Programme (Grant Agreement No. 101052200): projects No. ENR-MOD.01.MPG and ENR-MOD.03.NCSRD-01. Views and opinions expressed are however those of the author(s) only and do not necessarily reflect those of the European Union or the European Commission. Neither the European Union nor the European Commission can be held responsible for them. This work was supported in part by the Italian Ministry of Foreign Affairs and International Cooperation, grant number CN23GR02 and by the MMNLP project CSN4 of INFN, Italy.


\end{document}